\documentclass[preprint,showpacs,preprintnumbers,amsmath,amssymb]{revtex4}

\usepackage{graphicx}
\usepackage{dcolumn}
\usepackage{bm}
\usepackage{epsfig}

\begin{document}

\title{Exact results for the thermal and magnetic properties of strong coupling ladder compounds}

\author{Murray T. Batchelor$^1$, Xi-Wen Guan$^1$, Norman Oelkers$^1$, 
Kazumitsu Sakai$^2$, Zengo Tsuboi$^2$ and Angela Foerster$^3$}

\affiliation{$^1${Department of Theoretical Physics, RSPhysSE and Centre for Mathematics and its 
Applications, MSI,}\\
Australian National University, Canberra ACT 0200,  Australia\\
$^2$Institute for Solid State Physics, University of Tokyo, Kashiwanoha 5-1-5, Kashiwa, 
Chiba, 277-8581, Japan\\
$^3$Instituto de F\'{\i}sica da UFRGS,
                     Av.\ Bento Gon\c{c}alves, 9500,
                     Porto Alegre, 91501-970, Brasil}

\date{\today}

\begin{abstract}
\noindent
We investigate the thermal and magnetic properties of the integrable $su(4)$ ladder model 
by means of the quantum transfer matrix method.
The magnetic susceptibility, specific heat, magnetic entropy and high field magnetization 
are evaluated from the free energy derived via the recently proposed method of high 
temperature expansion for exactly solved models.
We show that the integrable model can be used to describe the physics of the strong coupling
ladder compounds.
Excellent agreement is seen between the theoretical results and the experimental data
for the known ladder compounds
($5$IAP)$_2$CuBr$_4$$\cdot$$2$H$_2$O, Cu$_{2}$(C$_5$H$_{12}$N$_2$)$_2$Cl$_4$ etc.

\end{abstract}

\pacs{75.10.Jm,64.40.Cn}

\keywords{Quantum spin ladders, integrable models}

\maketitle

The experimental realization of compounds with a ladder-like structure \cite{exp1} has
contributed to the intense interest in low-dimensional quantum systems.
The existence of a spin gap, magnetization plateaux, quantum critical points and 
superconductivity under hole doping are examples 
of key physical properties observed in the ladder compounds.
Of particular importance are the properties under a magnetic field $H$.
According to the perturbation theory result \cite{exp1,FT1}, 
the first-order terms for the zero temperature energy gap $\Delta $ 
and the critical field $H_{c2}$ are given in terms of the rung ($J_{\perp}$) and 
leg ($J_{\parallel}$) exchange couplings
by $\Delta=J_{\perp}-J_{\parallel}$ and $\mu_BgH_{c2}=J_{\perp}+2J_{\parallel}$.
These results are in good agreement with the experimental data for strong coupling ladder compounds.
However, the calculation of properties such as the full temperature phase diagram, the high field 
magnetization curve and the specific heat provide a significant challenge.

We demonstrate here that the integrable $su(4)$ ladder model \cite{ladd4,ladd3} is
capable of describing the physics of the ladder compounds.
Indeed, the thermodynamic Bethe Ansatz (TBA) applied to the integrable $su(4)$ ladder model
predicts the critical fields $ H_{c1}=J_{\perp}-4J_{\parallel}/\gamma$ and
$\mu_BgH_{c2}= J_{\perp}+4J_{\parallel}/\gamma$, where $\gamma$ is a rescaling parameter,
which are also good fits for the strong coupling compounds \cite{BGFZ}.
Very recently the high temperature expansion (HTE)  method  \cite{XXX,ZT} has 
suggested a way to calculate the full thermodynamic properties of integrable models 
from the so called $T$-system \cite{TS} appearing in the quantum transfer matrix (QTM) 
formalism \cite{QTM,note2}.
Here we extend this approach to the integrable ladder model to derive the thermal and
magnetic properties of the strong coupling ladder compounds.
We compare our results with the experimental data obtained for the compounds 
Cu$_2$(C$_5$H$_{12}$N$_2$)$_2$Cl$_4$ \cite{exp3, exp3-2} and
($5$IAP)$_2$CuBr$_4$$\cdot$$2$H$_2$O \cite{exp5}.

The Hamiltonian of the ladder model is \cite{ladd1,ladd4,ladd3,BGFZ}
\vspace{-5mm}
\begin{widetext}
\begin{equation}
{\cal H}=\frac{J_{\parallel}}{\gamma}\,{\cal H}_{{\rm leg}}+J_{\perp}\sum_{j=1}^{L}\vec{S}_j \vec{T}_j-
\mu_Bg H\sum_{j=1}^{L}(S_j^z+ T^z_j), \qquad 
{\cal H}_{{\rm leg}} = \sum_{j=1}^{L}\left(\vec{S}_j \vec{S}_{j+1}
+\vec{T}_j \vec{T}_{j+1}+ 4\, \vec{S}_j  \vec{S}_{j+1}  \vec{T}_j  \vec{T}_{j+1}\right),
\label{Ham}
\end{equation}
where $\vec{S }_j$ and $\vec{T}_j$ are Heisenberg operators, $\mu_B$ is the Bohr magneton and
$g$ is Land\'e factor.
Throughout, $L$ is the number of rungs and periodic boundary conditions are imposed.
In the strong coupling limit, the contribution to the low temperature physics from 
the multi-body term in ${\cal H}_{{\rm leg}} $ 
is minimal and, as a consequence, the integrable ladder Hamiltonian exhibits
similar critical behavior to the standard Heisenberg ladder \cite{BGFZ}.
We adapt the model into the QTM  method \cite{QTM}.
The eigenvalue of the QTM (up to a constant) is obtained by the nested Bethe Ansatz to be
\begin{eqnarray}
T^{(1)}_1(v,v^{(a)}_i)&=&e^{\beta \mu_1}\phi _-(v-\mathrm{i})
\phi _+(v)\frac{Q_1(v\!+\!\frac12{\mathrm{i}})}{Q_1(v\!-\!\frac12{\mathrm{i}})}
+e^{\beta \mu_2}\phi _-(v)\phi _+(v)
\frac{Q_1(v\!-\!\frac32{\mathrm{i}})Q_2(v)\ \ \ \,}{Q_1(v\!-\!\frac12{\mathrm{i}})Q_2(v\!-\!\mathrm{i})}\nonumber\\
& &+e^{\beta \mu_3}\phi _-(v)\phi _+(v)
\frac{Q_2(v\!-\!2\mathrm{i})Q_3(v\!-\!\frac12{\mathrm{i}})}{Q_2(v\!-\mathrm{i})
Q_3(v\!-\!\frac32{\mathrm{i}})}
+e^{\beta \mu_4}\phi _-(v)\phi _+(v+\mathrm{i})
\frac{Q_3(v\!-\!\frac52{\mathrm{i}})}{Q_3(v\!-\!\frac32{\mathrm{i}})}.\label{EQTM}
\end{eqnarray}
In this equation the chemical potential terms are $\mu_1\!\!=\!\!J_{\perp}/2$, 
$\mu_2 = \mu_B gH$, $\mu_3 =0$ and $\mu_4=- \mu_B gH$, 
with \mbox{$\phi _{\pm}(v)=(v\pm \mathrm{i}u_N)^{\frac{N}{2}}$}.
The inhomogeniety parameter $u_N\!=\!-\frac{J_{\parallel}\beta }{\gamma N}$, with \mbox{$Q_a(v)\!=\!\prod_{i=1}^{M^{(a)}}(v\!-\!v_i^{(a)})$},
for  $ a=1,2,3$. Here $N$ denotes  the Trotter-Suzuki number. 
The  fused  $T^{(a)}_m$ system \cite{TS}, which denotes the row-to-row transfer matrix with fusion type $(a,m)$ in the
auxiliary space carrying the $m$-fold symmetric tensor of the $a$-th fundamental
representation of the $su(4)$ algebra is essentially generated by the QTM eigenvalue $T^{(1)}_1$ in (\ref{EQTM}).
Thus $T^{(1)}_1$ 
can be embedded into the fused $T^{(a)}_m$ system.
The analytic non-zero and constant asymptotic properties of the normalized $\tilde{T}^{(a)}_m(v)$ system suggest the expansion ansatz
\begin{equation}
\lim_{N\rightarrow \infty}\tilde{T}_1^{(a)}(v)= {\rm exp} 
\left( \sum_{n=0}^{\infty}b_n^{(a)}(v)\left(\frac{J_{\parallel}}{\gamma T}\right)^n \right),\label{HTE}
\end{equation}
with $
b_n^{(a)}(v)=\sum^{n-1}_{j=0}c_{n,j}^{(a)}v^{2j}/(v^2+(a+1)^2/4)^n$.
The QTM eigenvalue satisfies a set of  the nonlinear integral equations \cite{ZT}
\begin{eqnarray}
\tilde{T}_1^{(a)}(v)&=&Q^{(a)}_1+\oint_{C_m^{(a)}}\frac{dy}{2\pi \mathrm{i}}\frac{1}{v-y-\beta_1^{(a)}}\left[\frac{\tilde{T}^{(a-1)}_1
(y+\beta_1^{(a)}-\frac12{\mathrm{i}})\tilde{T}^{(a+1)}_1(y+\beta_1^{(a)}-\frac12{\mathrm{i}})}{\tilde{T}^{(a)}_1(y+\beta_1^{(a)}-\mathrm{i})}\right]\nonumber\\
& &
+\oint_{\bar{C}_m^{(a)}}\frac{dy}{2\pi \mathrm{i}}\frac{1}{v-y+\beta_1^{(a)}}\left[\frac{\tilde{T}^{(a-1)}_1(y-\beta_1^{(a)}+\frac12{\mathrm{i}})\tilde{T}^{(a+1)}_1(y-\beta_1^{(a)}+
\frac12{\mathrm{i}})}{\tilde{T}^{(a)}_1(y-\beta_1^{(a)}+\mathrm{i})}\right],\,\,a=1,2,3. \label{nle}
\end{eqnarray}
Following Ref.~\cite{ZT}, the coefficients  $c_{n,j}^{(a)}$ can be obtained recursively 
from Eq.~(\ref{nle}) with initial conditions
$b_0^{(a)}=\ln Q_1^{(a)}$, where $Q^{(a)}_1$ are constants related to the chemical potential 
terms via $\lim_{N\to\infty}\lim_{|v|\to\infty}\tilde{T}^{(a)}_1(v)=Q^{(a)}_1$ with 
$Q^{(0)}_1=1$ and $Q^{(4)}_1=\exp(J_{\perp}/2T)$.
In this way the spin ladder free energy $f(T,H) =-T\ln T^{(1)}_1$ can be expanded in
powers of $J_{\parallel}/\gamma T$. 
For the first few orders we have
\begin{equation}
-\frac{1}{T} \, f(T,H)=\ln(2B_{\epsilon,1})  + A \left( \frac{J_{\parallel}}{\gamma T} \right) 
+\frac{3}{2}\left( A - A^2 + \frac{1}{2}\frac{\epsilon B_{1,\epsilon}}{B_{\epsilon,1}^3}\right)
\left(\frac{J_{\parallel}}{\gamma T}\right)^2 , \label{FE}
\end{equation}
\end{widetext}
where
$A = B_{\epsilon,0}(1+2B_{0,1})/B_{\epsilon,1}^2$
with $\epsilon=\exp(J_{\perp}/2T)$ and 
\begin{eqnarray}
B_{x,y}=x\cosh\left(\frac{J_{\perp}}{2T}\right)+y\cosh\left(\frac{\mu_BgH}{T}\right).
\end{eqnarray}
We find that the analytic expression (\ref{FE}) is sufficiently accurate to evaluate the model's thermodynamics.
Nevertheless, we have considered the HTE up to fifth order. 
%

\begin{center}
\begin{figure}[t]
\includegraphics[width=0.80\linewidth]{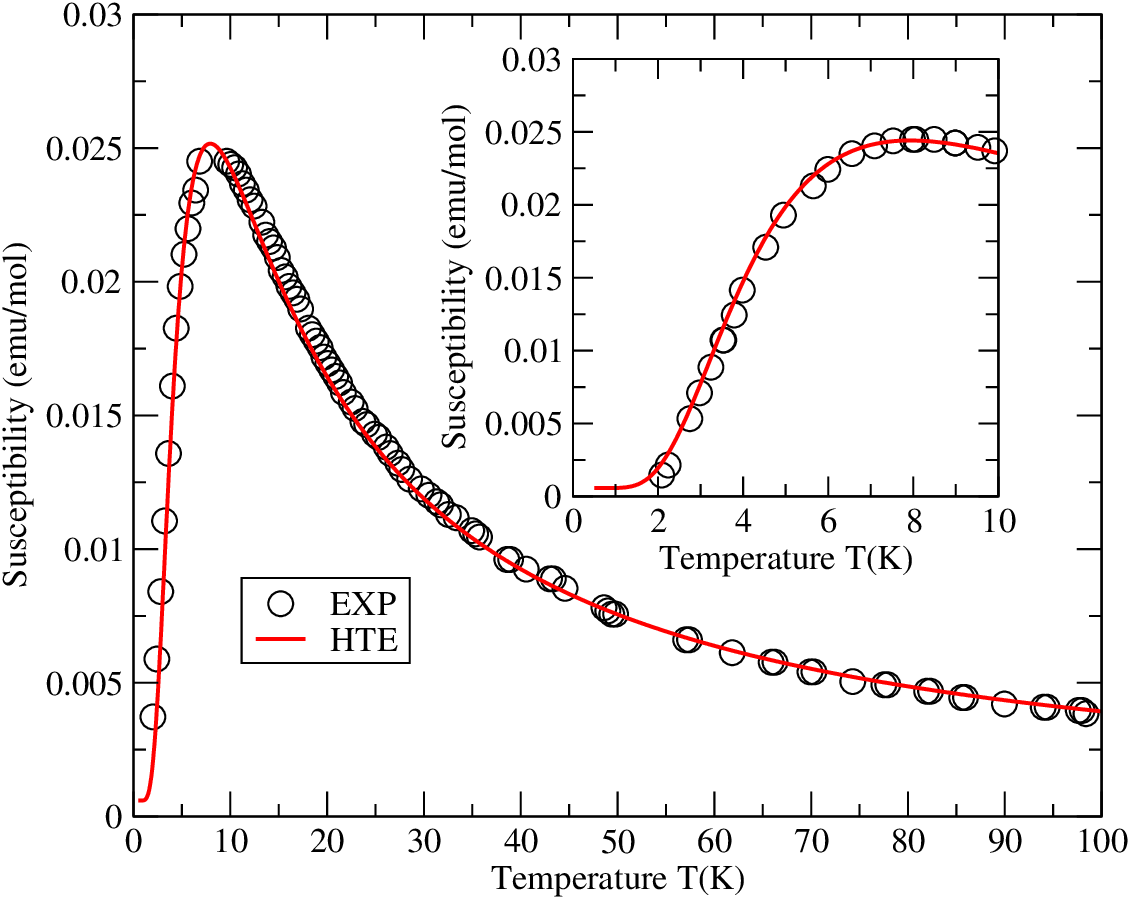}
\caption{
The susceptibility versus temperature for B$5$i$2$aT at $H=1$T \cite{exp5}. 
The solid line denotes the susceptibility evaluated directly from the HTE. 
A parameter fit suggests the coupling constants $J_{\perp}=13.3$K and $J_{\parallel}=1.15$K with $\gamma =4$,
$g=2.1$ and $\mu_B=0.672$K/T. 
The inset shows the same fit to the susceptibility at low temperature.}
\label{Fig1}
\end{figure}
\end{center}

The experimental measurements of the susceptibility and magnetization of the compound  
($5$IAP)$_2$CuBr$_4$$\cdot$$2$H$_2$O \cite{exp5} (abbreviated B$5$i$2$aT) suggest
a spin ladder with exchange couplings $J_{\perp}=13.0$K and $J_{\parallel}=1.15$K.
From the HTE for the integrable model we find that the values
$J_{\perp}=13.3$K and $J_{\parallel}=1.15$K with rescaling parameter
$\gamma =4$ give excellent fits to both the susceptibility and magnetization \cite{note1}.
The temperature dependence of the susceptibility is shown in Fig.~\ref{Fig1}.
The agreement with the theoretical curve derived from the HTE is clearly excellent.
The typical rounded peak for low magnetic field, characteristic of a
low dimensional antiferromagnet, is observed around  $8.1$K, in excellent agreement with
the experimentally estimated value of $8$K. 
%
%
The inset in Fig.~\ref{Fig1} shows  the low temperature behavior of the
susceptibility, which is in excellent agreement with the experimental data.

\begin{figure}[ht]
\includegraphics[width=0.80\linewidth]{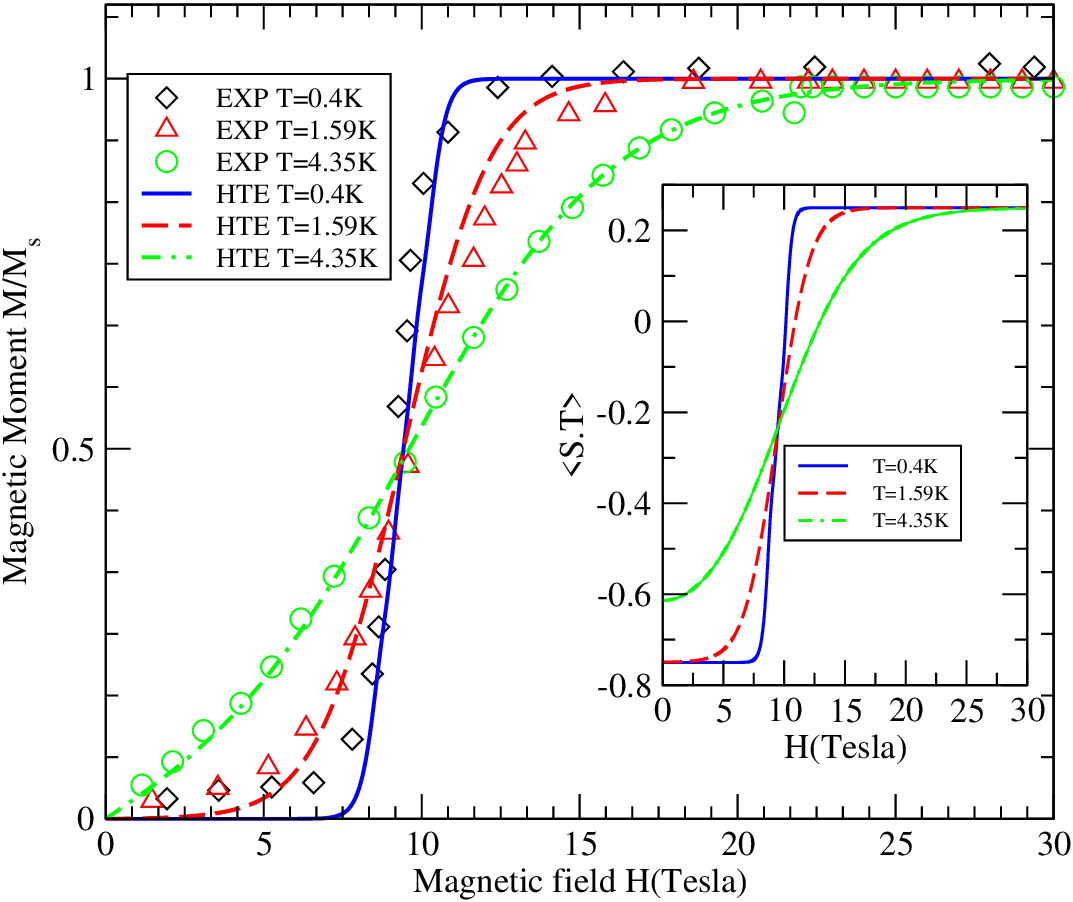}
\caption{Magnetization versus magnetic field for B$5$i$2$aT \cite{exp5} with the same constants as in 
Fig.~\ref{Fig1}.
The discrepancy in the magnetization curves at $T=0.4$K and $T=1.59 $K is due to paramagnetic impurities which 
become negligible for higher temperature. 
The inset shows the one-point correlation function vs magnetic field.}
\label{Fig2}
\end{figure}

The theoretical curves for the high field magnetization shown in Fig.~\ref{Fig2} for different temperatures 
are also in good agreement with the experimental values.
The field dependent magnetization curve predicts the low temperature phase diagram as well
as the magnetization plateaux. 
For very low temperature the rung singlet forms a dimerized groundstate
if the magnetic field is below the critical field $H_{c1}$. 
The length of the antiferromagnetic correlation is finite while the triplet state is gapfull.
For finite temperatures the triplet excitations are also involved in the gapped phase. 
This can be observed in the high field magnetization curves for $T=1.59 $K and $T=4.35 $K in Fig.~\ref{Fig2}.
At the critical field $H_{c1}$, the gap is closed with $\mu_B gH_{c1}=\Delta$. 
If the magnetic field is above the critical point $H_{c1}$, the lower triplet component becomes involved in the
groundstate. 
For zero temperature, it can be rigorously shown that the other two higher triplet components do not become involved 
in the groundstate \cite{BGFZ}. 
It follows that the strong coupling ladder can be mapped to the $XXZ$ Heisenberg chain with an effective magnetic field
term \cite{FT1,FT2}. 
The magnetization increases almost linearly with the field towards the critical point $H_{c2}$, where the ground state is 
fully polarized. 
At  $T=0.4$K, the HTE magnetization curve indicates $H_{c1} \approx 8.3$T and $H_{c2} \approx 10.5$T, which are in excellent
agreement with the experimental estimates of $8.4$T and $10.4$T.
The experimental magnetization in the singlet groundstate at low temperature appears to be nonzero.
This nonzero magnetization attributed to paramagnetic impurities, which drive the low-temperature deviation between the 
experimental and the theoretical curves in Fig.~\ref{Fig2}.

The inflection point is clearly visible in the experimental magnetization curves \cite{exp5}.
This point is also evident in the theoretical curves at $\mu_BgH \approx J_{\perp}$  where the
magnetization moment is $\frac12$.
The physical meaning of the inflection point is that the probabilities of the
singlet and the triplet states $|\!\!\uparrow \uparrow \rangle$ in the groundstate are equal.
Therefore, for the strong coupling ladder compounds at zero temperature 
the one-point-correlation function $\langle S_j\cdot T_j\rangle=-\frac{3}{4}$
lies in a gapped singlet groundstate, which indicates an ordered dimer phase, while
$\langle S_j \cdot T_j \rangle=\frac{1}{4}$ in the fully-polarized ferromagnetic phase.
However, in a Luttinger liquid phase, we find $\langle S_j\cdot T_j\rangle=-\frac{3}{4}+S^z$.
At low temperatures  $T\ll J_{\perp}$, the one point correlation function
is given by
$\langle S_j\cdot T_j\rangle=\frac14+\left(\frac{d}{dJ_{\perp}}f(T,H)\right)_T$.
The field-induced  quantum phase transitions can be clearly seen from the one-point correlation 
function curve shown in the inset of  Fig.~\ref{Fig2}.

\begin{center}
\begin{figure}[t]
\includegraphics[width=0.80\linewidth]{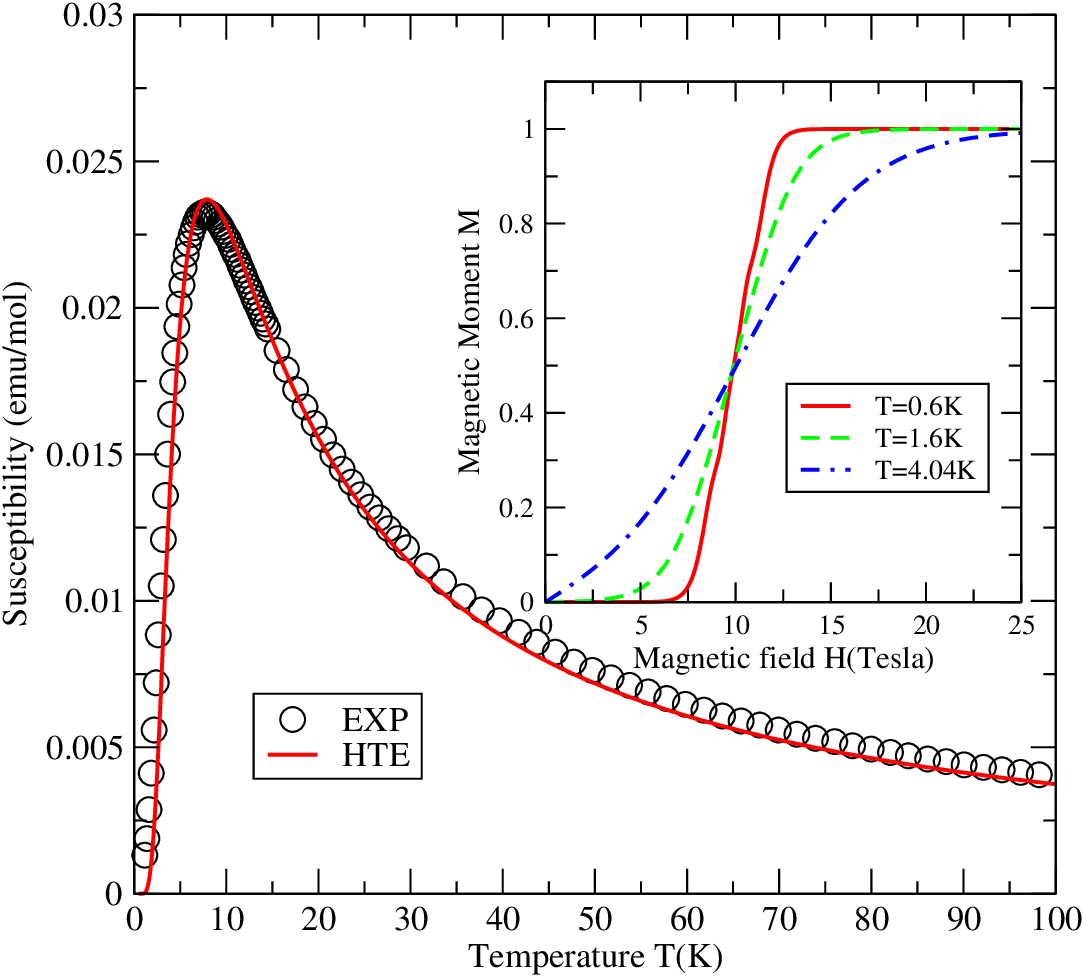}
\caption{Susceptibility versus temperature for the compound  Cu(Hp)Cl \cite{exp3-2}.
The solid line denotes the susceptibility evaluated directly from the HTE with $\mu_B=0.672$K/T,
$J_{\perp}=13.5$K, $J_{\parallel}=2.4$K, $\gamma =5$ and $g=2.03$.
The inset shows the magnetization versus magnetic field at different temperature. 
At $T=0.6$K, the critical fields are $H_{c1}\approx 7.8$T  and $H_{c2}\approx 13.0$T, 
in good agreement with the experimental results \cite{exp3,exp3-2}.}
\label{Fig3}
\end{figure}
\end{center}

We turn now to the ladder compound Cu$_2$(C$_5$H$_{12}$N$_2$)$_2$ Cl$_4$ \cite{exp3,exp3-2} 
(abbreviated Cu(Hp)Cl). 
In Fig.~\ref{Fig3}, we show the zero field magnetic susceptibility curve
obtained from the fifth order HTE free energy.
A  full fit with the experimental data suggests the coupling constants 
$J_{\perp}=13.5$K, $J_{\parallel}=2.4$K and $\gamma =5$.  
The effect of the magnetic field is to lift the susceptibility in the low temperature regime.
We notice that  there is a discrepancy with the zero temperature TBA result for $H_{c2}$  \cite{BGFZ}
due to the presence of  strong  exchange coupling along the legs. 
For finite temperature this discrepancy is smaller. 
The inset curves in Fig.~\ref{Fig3} show  the high field magnetization for temperatures $T=0.6, 1.6$ and $4.04$K. 
We observe that the critical points are $H_{c1}\approx 7.8$T and  $H_{c2}\approx 13.0 $T, which are in  good agreement 
with the experimental values \cite{exp3,exp3-2}.
It is also obvious that the finite temperature causes a spin-flip in the gapped ground state.

\begin{figure}[t]
\includegraphics[width=0.75\linewidth]{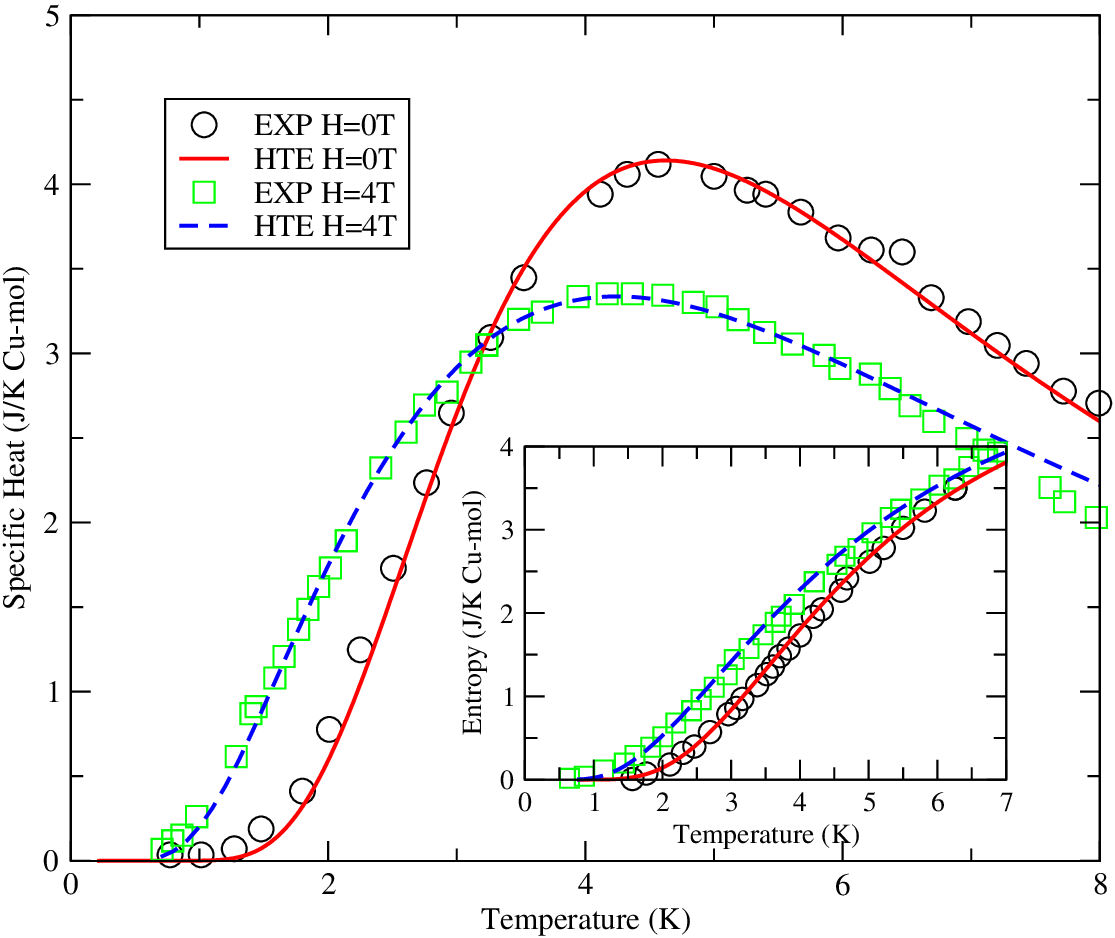}
\caption{Specific Heat  versus temperature for the compound Cu(Hp)Cl \cite{exp3-2} with  
the same constants as in Fig.~\ref{Fig3}.
The inset shows the field dependent entropy versus temperature. }
\label{Fig4}
\end{figure}

The  specific heat curves in Fig.~\ref{Fig4} for $H=0$T and $H=4$T
indicate that the HTE result also agrees satisfactorily with the experimental data \cite{exp3-2}.
In the absence of a magnetic field, a rounded peak indicating short range ordering is observed around $4.5$K. 
For temperatures $T<4.5$K, there is an exponential decay due to an ordered phase.
The humps become smaller as the magnetic field increases. 
For the $H=4$T curve  a peak is observed at around $4$K. 
As to be expected, there appears to be a small deviation from  the experimental data  at very low temperatures. 
The inset of  Fig.~\ref{Fig4} shows that  the entropy curves for magnetic fields $H=0$T and $H=4$T
are also in agreement with the experimental data \cite{exp3-2}.

\begin{figure}[t]
\vspace{3mm}
\includegraphics[width=0.75\linewidth]{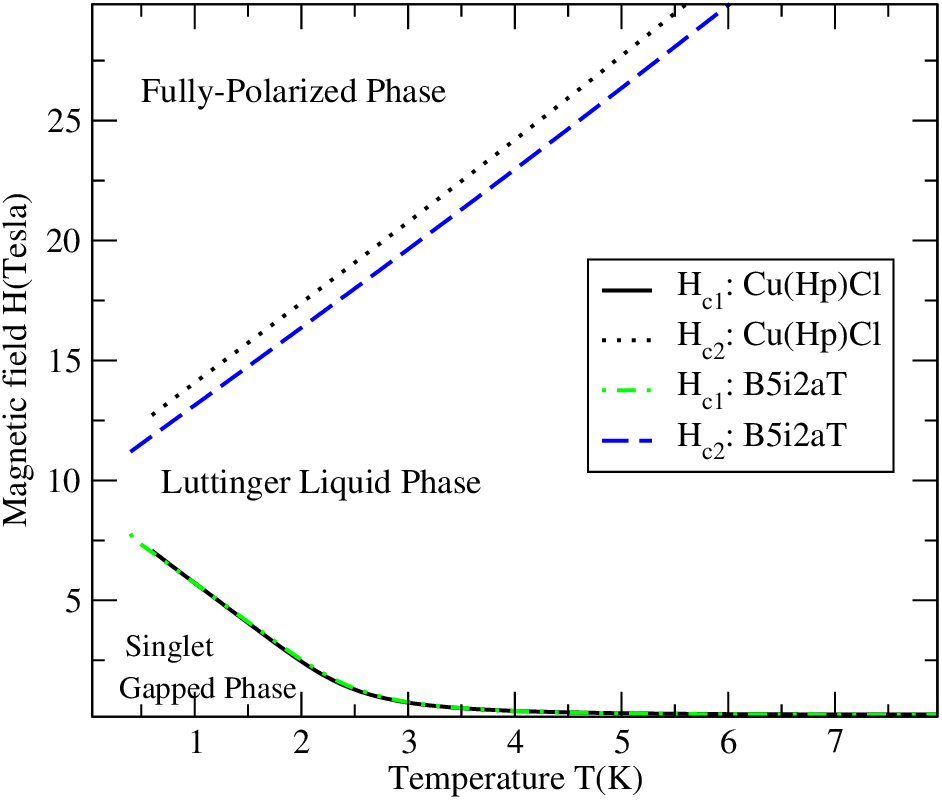}
\caption{
Phase diagrams  for the compounds B$5$i$2$aT and Cu(Hp)Cl.
%
}
\label{Fig5}
\end{figure}
The full phase diagram for the two compounds is shown in Fig.~\ref{Fig5}.
The slopes of the critical curves indicate  that  the estimated values of $H_{c1}$ and $H_{c2}$ at 
$T=0$K coincide with the TBA results \cite{BGFZ}. 
We have also examined other strong coupling compounds.
Comparison with the experimental data for the compound BIP-BNO \cite{BIP}
suggests the coupling constants $J_{\perp}=75$K, $J_{\parallel}=15$K with 
$\gamma =6.0$ and $g=2.0$.
For the compound [Cu$_2$(C$_2$O$_2$)(C$_{10}$H$_8$N$_2$)$_2$)](NO$_3$)$_2$ \cite{SC2}
we find $J_{\perp}=515$K, $J_{\parallel}=40$K with $\gamma =5.0$ and $g=2.14$.
The respective spin excitation gaps, $\Delta \approx 52 $K and $\Delta\approx 460 $K, are also 
in good agreement with the experimental values. 
The application of the HTE method to other ladder models, such as the mixed-spin ladders 
should now be straightforward. 
We also note that the HTE method predicts a fractional magnetization plateau with
respect to different Land\'{e} $g$-factors in the $su(4)$ spin-orbital model, 
which coincides with the Bethe ansatz \cite{SO1} and TBA \cite{SO2} results.

\noindent
{\em Acknowledgements.} This work has been supported by the Australian Research Council. 
KS is supported in part by a JSPS Research Fellowship for Young Scientists.
AF thanks FAPERGS and CNPq.
We thank H.-Q. Zhou and Z.-J. Ying for helpful discussions.

\begin{thebibliography}{99}

\bibitem{exp1} For reviews, see E. Dagotto and T.M. Rice, Science 271 (1996) 618; 
E. Dagotto, Rep. Prog. Phys. 62 (1999) 1525.

\bibitem{FT1} M. Reigrotzki, H. Tsunetsugu and T.M. Rice, J. Phys. Condens. Matter 6 (1994) 9235.

\bibitem{ladd4} Y. Wang, Phys. Rev. B 60 (1999) 9236.

\bibitem{ladd3} M.T. Batchelor and M. Maslen, J. Phys. A 32 (1999) L377; 
M.T. Batchelor, J. de Gier and M. Maslen, J. Stat. Phys. 102 (2001) 559.

\bibitem{BGFZ} M.T. Batchelor, X.-W. Guan, A. Foerster and  H.-Q. Zhou, 
New. J. Phys. 5 (2003) 107.

\bibitem{XXX} M. Shiroishi and M. Takahashi, Phys. Rev. Lett. 89 (2002) 117201.

\bibitem{ZT} Z. Tsuboi, J. Phys. A 36 (2003) 1493.

\bibitem{TS} A. Kuniba, T. Nakanishi and J. Suzuki, Int. J. Mod. Phys. A 9 (1994) 5215.

\bibitem{QTM} M. Suzuki, Phys. Rev. B 31 (1985) 2957; 
A.Kl\"{u}mper, Ann. Physik 1 (1992) 540;
G. J\"{u}ttner, A. Kl\"{u}mper and J. Suzuki, Nucl. Phys. B 487 (1997) 650.

\bibitem{note2} A different QTM method has been applied numerically to the standard
Heisenberg ladder in M. Troyer, H. Tsunetsugu and D. W\"urtz, 
Phys. Rev. B 50 (1994) 13515.

\bibitem{exp3} G. Chaboussant et al., Phys. Rev. B 55 (1997) 3046;
Phys. Rev. Lett. 79 (1997) 925;
Phys. Rev. Lett. 80 (1998) 2713.

\bibitem{exp3-2} M. Hagiwara, H.A. Katori, U. Schollw\"{o}ck and H.-J. Mikeska,
Phys. Rev. B 62 (2000) 1051.

\bibitem{exp5} C.P. Landee, M.M. Turnbull, C. Galeriu, J. Giantsidis and F.M. Woodward, 
Phys. Rev. B 63 (2001) 100402.

\bibitem{ladd1} The multi-body term arises naturally in other contexts, see, e.g., 
A.A. Nersesyan and A.M. Tsvelik, Phys. Rev. Lett. 78
(1997) 3939; A.K. Kolezhuk and H.-J Mikeska, Phys. Rev. Lett. 80 (1998) 2709.

\bibitem{note1}
An overall unit constant in the susceptibility needs to be chosen to  match the experimental scaling as the $T$-system is only determined up to a multiplicative factor. 

\bibitem{FT2} G. Chaboussant et al., Eur. Phys. J. B 6 (1998) 167. 
\bibitem{BIP}K. Katoh, Y. Hosokoshi, K. Inoue and T. Goto, J. Phys. Soc. Japan 69 (2000) 1008.
\bibitem{SC2}Z. Honda, Y. Nonomura and K. Katsumata, J. Phys. Soc. Japan 66 (1997) 3689.

\bibitem{SO1} S.-J. Gu, Y.-Q. Li and H.-Q. Zhou, cond-mat/0308432.
\bibitem{SO2}
 Z.-J. Ying, A. Foerster, X.-W. Guan, B. Chen and I. Roditi, cond-mat/0308443.

\end{thebibliography}

\end{document}